\documentclass[preprintnumbers,superscriptaddress,showkeys,showpacs,byrevtex]{revtex4}
\usepackage{amsmath,amsfonts,amssymb,amscd,amsxtra,amsthm}
\usepackage{graphicx}
\usepackage{bm}
\usepackage{epstopdf}
\usepackage{multirow}
\begin{document}
\preprint{KIAS-P11068}
\preprint{CYCU-HEP-11-21}
\title{Fragmentation functions and parton distribution functions\\
 for the pion with the nonlocal interactions}
\author{Seung-il Nam}
\email[E-mail: ]{sinam@kias.re.kr}
\affiliation{School of Physics, Korea Institute for Advanced Study (KIAS), Seoul 130-722, Korea}
\author{Chung-Wen Kao}
\email[E-mail: ]{cwkao@cycu.edu.tw}
\affiliation{Department of Physics, Chung-Yuan Christian University (CYCU), Chung-Li 32023, Taiwan}
\date{\today}
\begin{abstract}
We study the unpolarized fragmentation functions and parton distribution functions of the pion employing the nonlocal chiral quark model.  This model manifests the nonlocal interactions between the quarks and pseudoscalar mesons in the light-cone coordinate. It turns out that the nonlocal interactions result in substantial differences in comparison to typical models with only local couplings. We also perform the high $Q^2$-evolution for our results calculated at a relatively low renormalization scale $Q^2\approx0.36\,\mathrm{GeV}^2$ to be compared with the experimental data. Our results after evolution are in qualitatively good agreement with those data.
\end{abstract}
\maketitle
\section{Introduction}

To apply perturbative Quantum Chromodynamics (QCD) to study hadronic processes, such as the deep-inelastic electron-proton scattering, one needs the QCD factorization theorem to guarantee the cross section to be the convolution of the two parts: the process-dependent perturbative QCD (pQCD) calculable short-distance parton cross section, and the universal long-distance functions which can be extracted from experiments. Since the strong interaction in long-distance is essentially nonperturbative, those long-distance functions are incalculable by pQCD.

To date, there has been increasing interest in those long-distance functions such as the fragmentation functions and parton distribution functions. The fragmentation function characterizes the probability for a hadron fragmented from a quark with the momentum fraction $z$. It plays a crucial role in analyzing the semi-inclusive processes in the electron-positron scattering, deep-inelastic proton-proton scattering, and so on~\cite{Collins:1992kk,Mulders:1995dh,Boer:1997nt,Anselmino:1994tv,Anselmino:2008jk,Christova:2006qs,Anselmino:2007fs,Bacchetta:2006tn,Efremov:2006qm,Collins:2005ie,Ji:2004wu}. The parton distribution function provides information for the distribution of the momentum fraction $x$ carried by a parton inside a hadron. It is a necessary ingredient in studying hard scattering processes such as the deep-inelastic electron-proton scattering. It is worth mentioning that a parton distribution function can be extracted from the corresponding fragmentation function via the Drell-Levy-Yan (DLY) relation due to the analytical continuation~\cite{Drell:1969jm}. { This analytic continuation is possible only if the parton distribution function and the fragmentation function are assumed to be described by the same function defined in the different regions~\cite{Drell:1969jm}.}

Because of these salient physical meanings implied in those functions, they have been intensively studied so far~\cite{Kretzer:2000yf,Conway:1989fs,deFlorian:2007aj,Sutton:1991ay,Hirai:2007cx,Kniehl:2000fe,Ji:1993qx,Jakob:1997wg, Bacchetta:2002tk,Amrath:2005gv,Bacchetta:2007wc,Meissner:2010cc,Bentz:1999gx,Ito:2009zc,Matevosyan:2010hh,Matevosyan:2011ey,Matevosyan:2011vj,Nguyen:2011jy,
Aloisio:2003xj,Dorokhov:1991nj,Nam:2006sx,Nam:2006au,Praszalowicz:2001pi,Noguera:2005cc,Aicher:2010cb,Holt:2010vj}. These functions have been extracted from the available empirical data by global analyses or from the parametrizations which satisfy certain constraints~\cite{Kretzer:2000yf,Conway:1989fs,deFlorian:2007aj,Sutton:1991ay,Hirai:2007cx,Kniehl:2000fe}. Many theoretical investigations on the fragmentation functions have been carried out. For example, a rigorous QCD proof has done for the momentum sum rules of the fragmentation functions~\cite{Meissner:2010cc}. The Dyson-Schwinger (DS) method was applied to compute the valance-quark distribution function, resulting in relatively good agreement with the available experimental data for the pion and kaon~\cite{Nguyen:2011jy}. Similarly, the pion parton distribution was studied with a nonlocal Lagrangian using the DS method~\cite{Noguera:2005cc}. In Ref.~\cite{Aicher:2010cb}, the valance-quark distribution was determined by analyzing the Drell-Yan process including the soft-gluon resummation. A review for the experimental and theoretical status for valance-quark distribution for the nucleon and pion is given in Ref.~\cite{Holt:2010vj}. The authors of Refs.~\cite{Dorokhov:1991nj,Nam:2006sx,Nam:2006au,Praszalowicz:2001pi} have also made use of the instanton-motivated approaches to compute the quark distribution functions as well. The collins fragmentation functions which play an important role in the transverse spin physics have also been studied ~\cite{Bacchetta:2002tk,Amrath:2005gv,Bacchetta:2007wc}. Monte-Carlo simulations with supersymmetric (SUSY) QCD were carried out to obtain the fragmentation function up to a very high energy in the center-of-mass frame $\sqrt{s}$~\cite{Aloisio:2003xj}.

Among those approaches, we are in particular interested in the ones based on the chiral dynamics. Because the strong interaction in long-distance is dominated by the chiral physics, therefore the nonperturbative objects such as fragmentation functions also should be subjected to
the chiral physics. Along this line, the fragmentation functions were first studied by Georgi-Manohar model in Refs.~\cite{Collins:1992kk, Ji:1993qx}, and later by chiral-quark-meson coupling models in the pseudoscalar (PS) and pseudovector (PV) schemes~\cite{Jakob:1997wg, Bacchetta:2002tk,Amrath:2005gv}. Their results indicate considerable differences between these two schemes. In Refs.~\cite{Bentz:1999gx,Ito:2009zc,Matevosyan:2010hh,Matevosyan:2011ey,Matevosyan:2011vj}, the authors studied the fragmentation functions for various hadrons in terms of the Nambu-Jona-Lasinio (NJL) model. In their works, quark-jets and resonances were taken into account and consequently the momentum sum rules is satisfied. The quark-jet contributions turned out to be crucial to reproduce the various fragmentation functions for the small $z$ region. Besides, the quark-jet is also essential to generate the unfavored fragmentation functions~\cite{Ito:2009zc,Matevosyan:2011ey,Matevosyan:2011vj}.

In the present work, we concentrate on the fragmentation functions and quark distribution functions for the SU(2) light flavor for the positively charged pion, i.e. $\pi^+\equiv u\bar{d}$. We only evaluate the $\pi^+$ case because the other isospin channels can be easily obtain by multiplying the corresponding isospin factors {in our model}~\cite{Matevosyan:2011ey}. For this purpose, we employ the nonlocal chiral quark model (NLChQM). This model is based on the dilute instanton-liquid model (LIM), {which is properly defined in Euclidean space}~\cite{Shuryak:1981ff,Diakonov:1985eg,Diakonov:1983hh,Schafer:1996wv,Diakonov:2002fq}. It is worth mentioning that, NLChQM have successfully produced results for various {nonperturbative QCD quantities} which are in agreement with experiments as well as lattice QCD (LQCD) simulations.~\cite{Musakhanov:1998wp,Musakhanov:2002vu,Nam:2007gf,Nam:2010pt,Dorokhov:2000gu,Dorokhov:2002iu}.

Our calculations are performed in the light-cone coordinate at a relatively low renormalization scale $\mu\approx0.6$ GeV, corresponding to the two phenomenological instanton parameters, i.e. average size of the (anti)instanton $\bar{\rho}\approx1/3$ fm and average inter-instanton distance $\bar{R}\approx1$ fm for the dilute (anti)instanton ensemble. These instanton parameters have been determined phenomenologically~\cite{Shuryak:1981ff} or generated by LQCD simulations ~\cite{Negele:1998ev}. Although the fragmentation functions and the parton distribution functions are defined properly only in Minkowski space and there has been no rigorous proof, we still assume that there is an appropriate analytic continuation between the instanton physics and the NLChQM defined in the light-cone coordinate. We notice that there have been successful applications based on this assumption to various physical quantities~\cite{Dorokhov:1991nj,Nam:2006sx,Nam:2006au,Praszalowicz:2001pi}.

Distinctive features of the present approach, in comparison to other {\it local} effective chiral models such as the conventional NJL model and usual quark-meson coupling models, are as follows: 1) The couplings between quarks and pseudoscalar (PS) mesons are nonlocal. Those nonlocal interactions are led by the intricate quark-(anti)instanton interactions inside the instanton ensemble, flipping quark helicities~\cite{Shuryak:1981ff,Diakonov:1985eg,Diakonov:1983hh,Schafer:1996wv,Diakonov:2002fq}. 2) As a result, the quark-PS meson coupling, which is represented by the momentum-dependent constituent-quark mass, plays the role of the natural UV regulator. Hence, one does not need any artificial regulators. Interestingly enough, the computed momentum-dependence of the mass turns out to be very comparable with the LQCD simulations~\cite{Diakonov:2002fq,Bowman:2002kn}. 3) Moreover, all the relevant physical quantities such as the low-energy constants (LECs) are determined self-consistently only with the instanton parameters~\cite{Goeke:2007bj,Musakhanov:2008tq,Goeke:2010hm}. There is no more adjustable parameters in principle within the framework as long as we are considering the light-flavor SU(2) sector. We emphasize that, among the above features, the nonlocality of the interaction is the most important ingredient which has never appeared in other chiral models.

As for the numerical calculations, we compute the {\it elementary} fragmentation function $d^\pi_u(z)$, standing for the fragmentation process $u\to\pi q'$, where $\pi$ is assigned for $\pi^+$ throughout this work, i.e. the quark $u$ quark is fragmented into the pion and an energetic quark $q'$. Since an energetic quark will be fragmented into other hadrons until it possesses insufficient energy to be fragmented into hadrons, this elementary fragmentation function does not satisfy the momentum sum rule~\cite{Ito:2009zc,Matevosyan:2011ey}. It is likely that if the quark jets and resonance contributions are taken into account, the sum rule will be satisfied~\cite{Ito:2009zc,Matevosyan:2011ey}. However we will not consider those effects for brevity in the present work. Our results of the fragmentation functions turn out to be substantially larger in comparison to other models.
Actually the first moment of the elementary fragmentation function can be express as~\cite{Ito:2009zc}
\begin{equation}
\int^{1}_{0} dz \sum_{\pi} d^{\pi}_{q}(z)=1-Z_{Q}.
\end{equation}
Here $Z_{Q}$ is the residue of the quark propagator in the presence of the pion cloud and can be interpreted as the probability of finding
a bare constituent quark without the pion cloud. In other words, those elementary fragmentation functions are normalized to the number of pions per quark~\cite{Ito:2009zc}. The value of $Z_{Q}$ in our model is much smaller compared with other models with local couplings. Physically we can understand the difference between our model and other models is due to that the nonlocal couplings between quarks and PS mesons in our model somehow mimic {a} part of the meson-cloud effects. Therefore our {\it elementary} fragmentation functions have already contained more pion-cloud contributions than other models. To investigate the $z$-dependence of the fragmentation functions from different models, we renormalize these elementary fragmentation functions as follows
\begin{equation}
\int^{1}_{0} dz \sum_{\pi} D^{\pi}_{q}(z)=1.
\end{equation}
We find that our fragmentation functions multiplied by $z$, i.e. $zD^\pi_u$ is more symmetric with respect to $z$
compared with the ones from other models.

The calculated fragmentation functions are evolved to high $Q^2$ by the DGLAP evolution using the code {\it QCDNUM}~\cite{Botje:2010ay,DGLAP} to compare with the empirical data. By doing so, we see qualitative agreement between our result and the empirical data, although some underestimates are shown in the small-$z$ region. This defect is expected to be improved by the inclusion of the quark-jet contributions~\cite{Ito:2009zc,Matevosyan:2011ey}.

The parton distribution function $f^\pi_u(x)$ is {extracted} from the fragmentation function by the DLY relation. As done in Ref.~\cite{Ito:2009zc}, we provide the numerical results for the minus-type and plus-type quark distribution functions {being multiplied by $x$,} i.e. $x(f^\pi_u-f^\pi_{\bar{u}})$ and $x(f^\pi_u+f^\pi_{\bar{u}})$, respectively. Note that the minus-type one is nothing but the valance quark-distribution function. Again, we perform the high $Q^2$ evolution in order to compare our results with the empirical ones, resulting in qualitative agreement except the sizable underestimate observed again in the small-$x$ region after the evolution up to $Q^2=4\,\mathrm{GeV}^2$ for the minus-type one. Again, this discrepancy implies more realistic treatments beyond the elementary process, such as the quark-jet contributions as mentioned previously~\cite{Ito:2009zc}, is required. However, implanting the quark-jet contributions into the present nonlocal-interaction models is challenging, since the momentum-dependent quark-PS meson coupling can provide nontrivial effects to the DLY relation and the quark-jet contributions here.
So we leave it for our future study.
It also turns out that the high $Q^2$ evolution up to $Q^2=27\,\mathrm{GeV}^2$ reproduces the experimental data qualitatively well.

The present report is structured as follows: In Section II, we briefly introduce the model we use and sketch our computation procedure. In Section III we present our numerical results and related discussions. The final Section is devoted for the summary, conclusion, and future perspectives.
\section{Nonlocal pion-quark coupling}
\begin{figure}[t]
\includegraphics[width=12cm]{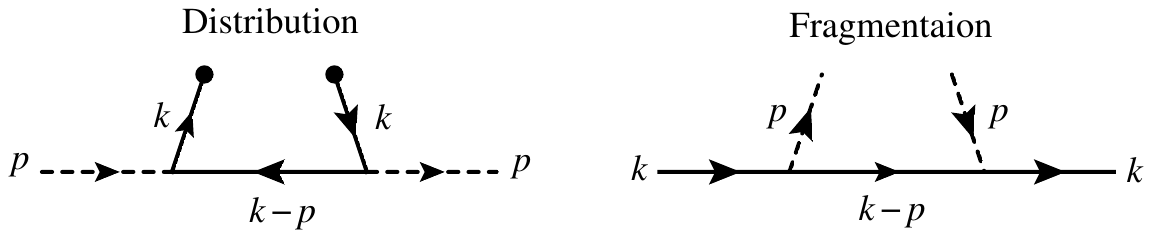}
\caption{Schematic figures for the quark-distribution (left) and fragmentation (right) functions, in which the solid and dash lines denote the quark and pseudoscalar meson, respectively.}
\label{FIG0}
\end{figure}
The unpolarized fragmentation function $D^h_{q}$ indicates the process that an off-shell quark ($q$) is fragmented into {an unobserved set of particles} ($X$) and on-shell hadron ($h$), i.e. $q\to h\,X$. A schematic figure for this function is given in the right of Figure~\ref{FIG0}. Here, the fragmentation function is defined in the light-cone coordinate, assuming the light-cone gauge, as follows~\cite{Bacchetta:2002tk,Amrath:2005gv}:
\begin{equation}
\label{eq:FRAG}
D^h_{q}(z,\bm{k}^2_T,\mu)=\frac{1}{4z}\int dk^+\mathrm{Tr}
\left[\Delta(k,p,\mu)\gamma^- \right]|_{zk^-=p^-},
\end{equation}
where $k$, $p$, and $z$ stand for the four-momenta of the initial quark and fragmented hadron, and the longitudinal momentum fraction possessed by the hadron, respectively. $k_{\pm}$ denotes $(k_{0}\pm k_{3})/\sqrt{2}$ in the light-cone coordinate. All the calculations are carried out in the frame where $\bm{k}_{\perp}=0$. Here the $z$-axis is chosen to be the direction of $\vec{k}$. On the other hand, $\bm{k}_{T}=\bm{k}-\left(\frac{\bm{k}\cdot\bm{p}}{|\bm{p}|^2}\right)\bm{p}$, defined as the
the transverse momentum of the initial quark with respect to the direction of the momentum of the produced hadron, is nonzero.
$\mu$ denotes the renormalization scale at which the fragmentation process computed. Note that we consider this renormalization scale is almost the same with the momentum-transfer scale, i.e. $\mu^2\approx Q^2$ for simplicity, unless otherwise stated. The correlation $\Delta(k,p,\mu)$ reads generically:
\begin{equation}
\label{eq:COR1}
\Delta(k,p,\mu)=\sum_X\int\frac{d^4\xi}{(2\pi)^4}e^{+ik\cdot\xi}
\langle0|\psi(\xi)|h,X\rangle\langle h,X|\psi(0)|0\rangle.
\end{equation}
Here $\psi$ represents the quark field, whereas $\xi$ the spatial interval on the light cone.
Furthermore one can integrate over $\bm{k}_{T}$,
\begin{equation}
D^h_{q}(z,\mu)=\pi z^2\int^{\infty}_{0}d\bm{k}^{2}_T\,D^{h}_{q}(z,\bm{k}_{T},\mu).
\end{equation}
The factor of $z^2$ is due to the fact that the integration is over  
$\bm{p}_\perp=\bm{p}-\left(\frac{\bm{p}\cdot \bm{k}}{|\bm{k}|^2}\right)\bm{k}$,
the transverse momentum of the produced hadron
with respect to the quark direction, and there is a relation between $\bm{p}_{\perp}$ and $\bm{k}_{T}$: 
$\bm{p}_\perp=-z\bm{k}_{T}$
The integrated fragmentation function satisfies the momentum sum rule:
\begin{equation}
\label{eq:SUM}
\int^1_0\sum_{h}zD^h_{q}(z,\mu)\,dz=1,
\end{equation}
where $h$ indicates for all the possible hadrons fragmented. Eq.~(\ref{eq:SUM}) means that all of the momentum of the initial quark $q$ is transferred into the momenta of the fragmented hadrons. From the Drell-Levi-Yan (DLY) relation~\cite{Drell:1969jm,Amrath:2005gv,Bentz:1999gx,Ito:2009zc,Matevosyan:2011ey}, $D^h_{q}$ can be related to the parton distribution function $f^h_{q}$, provided that there is a proper analytic continuation. The relation is as follows,
\begin{equation}
\label{eq:DLY}
D^h_{q}(z)=\frac{z}{6}f^h_{q}\left(x\right),\,\,\,\,\mathrm{where}\,\,\,\,x=\frac{1}{z},
\end{equation}
where $x$ denotes the momentum fraction possessed by a parton inside the hadron. A schematic figure for the quark-distribution function is depicted in the left of Figure~\ref{FIG0}.

In this article, we use NLChQM to investigate these nonperturbative objects, i.e. fragmentation and parton distribution functions. This model is motivated from the dilute instanton liquid model~\cite{Diakonov:1985eg,Shuryak:1981ff,Diakonov:1983hh,Diakonov:2002fq,Schafer:1996wv}. We note that, to date, various nonperturbative QCD properties have been well studied in terms of the instanton vacuum configuration and the results are comparable with experiments and LQCD simulations~\cite{Nam:2007gf,Nam:2010pt,Musakhanov:1998wp,Musakhanov:2002vu}. In that model, nonperturbative QCD effects are deciphered by the nontrivial quark-instanton interactions in the dilute instanton ensemble. However this model by nature is defined in Euclidean space because the (anti)instantons are well defined there by signaling the tunneling between the infinitely degenerate QCD vacua. Although there have been no rigorous derivation on the analytic continuation from the instanton physics to those in Minkowski one, there are still several challenging studies which try to apply the idea of the instanton physics to the physical quantities defined properly only in Minkowski space, such as the light-cone wave function~\cite{Dorokhov:1991nj,Nam:2006sx,Nam:2006au,Praszalowicz:2001pi}. Following those studies we adopt the effective chiral action (EChA) from NLChQM in Minkowski space as follows:
\begin{equation}
\label{eq:ECA}
\mathcal{S}_\mathrm{eff}[m_f,\phi]=-\mathrm{Sp}\ln\left[i\rlap{/}{\partial}
-\hat{m}_f-\sqrt{M(\loarrow{\partial}^2)}U^{\gamma_5}\sqrt{M(\roarrow{\partial}^2)}\right],
\end{equation}
where $\mathrm{Sp}$ and $\hat{m}_f$ denote the functional trace $\mathrm{Tr}\int d^4x \langle x|\cdots|x\rangle$ over all the relevant spin spaces and SU(2) current-quark mass matrix $\mathrm{diag}(m_u,m_d)$, respectively. Throughout the present work, we will take $m_u=m_d=5$ MeV, {considering the isospin space}. Note that, in deriving EChA in Eq.~(\ref{eq:ECA}), we simply replace the Euclidean metric for the (anti)instanton effective chiral action into that for Minkowski space. The momentum-dependent effective quark mass generated from the interactions between the quarks and nonperturbative QCD vacuum, can be written in a simple $n$-pole type form factor as follows~\cite{Dorokhov:1991nj,Nam:2006sx,Nam:2006au,Praszalowicz:2001pi}:
\begin{equation}
\label{eq:MDM}
M(\partial^2)=M_0\left[\frac{n\Lambda^2}{n\Lambda^2-\partial^2+i\epsilon} \right]^{n},
\end{equation}
where $n$ indicates a positive integer number. We will take $n=2$ as in the instanton model~\cite{Nam:2006sx,Nam:2006au}. $\Lambda$ stands for the model renormalization scale. It is related to the average (anti)instanton size $\bar{\rho}$. The nonlinear PS-meson field, i.e. $U^{\gamma_5}$ takes a simple form~\cite{Diakonov:2002fq} with the normalization following Ref.~\cite{Bacchetta:2002tk} to be consistent with the definition of the fragmentation function in Eq.~(\ref{eq:FRAG}):
\begin{equation}
\label{eq:CHIRALFIELD}
U^{\gamma_5}(\phi)=
\exp\left[\frac{i\gamma_{5}(\bm{\tau}\cdot\bm{\phi})}{2F_{\phi}}\right]
=1+\frac{i\gamma_{5}(\bm{\tau}\cdot\bm{\phi})}{2F_{\phi}}
-\frac{(\bm{\tau}\cdot\bm{\phi})^{2}}{8F^{2}_{\phi}}+\cdots,
\end{equation}
where $F_\phi$ stands for the weak-decay constant for the PS meson $\phi$. For instance, $F_\pi$ is chosen to be about $93$ MeV in this normalization. {We note that, however, the value of $F_\pi$ can be determined rather phenomenologically to reproduce relevant physical quantities and conditions, even in NLChQM, and will discuss this in detail in Section III. We also write explicitly the PS-meson fields:
\begin{equation}
\label{eq:PHI}
\bm{\tau}\cdot\bm{\phi}=\left(
\begin{array}{cc}
\pi^{0}&\sqrt{2}\pi^{+}\\
\sqrt{2}\pi^-&-\pi^{0}\\
\end{array} \right).
\end{equation}
By expanding the nonlinear PS-meson field up to $\mathcal{O}(\phi^1)$ from EChA in Eq.~(\ref{eq:ECA}), one can derive the following effective interaction Lagrangian density {in the coordinate space} for the nonlocal quark-quark-PS meson vertex:
\begin{equation}
\label{eq:LAG}
\mathcal{L}_{\phi qq}=\frac{i}{2F_\phi}\bar{q}\sqrt{M(\loarrow{\partial}^2)}
\gamma_5(\bm{\tau}\cdot\bm{\phi})\sqrt{M(\roarrow{\partial}^2)}q.
\end{equation}

Then, the correlation in Eq.~(\ref{eq:COR1}) can be evaluated using this effective Lagrangian for the {\it elementary} fragmentation function $q(k)\to\phi(p)+\,q'(r)$, {where $r=k-p$}:
\begin{eqnarray}
\label{eq:DELTA}
\Delta(k,p,\mu)&=&-\frac{1}{(2\pi)^4}\frac{M_kM_r}{2F^2_\phi}
\left(\frac{\rlap{/}{k}+\bar{M}_k}{k^2-\bar{M}^2_k}\right)\gamma_5
\left(\rlap{/}{r}+\bar{M}_r\right)\gamma_5
\left(\frac{\rlap{/}{k}+\bar{M}_k}{k^2-\bar{M}^2_k}\right)
\left[2\pi\delta(r^2-\bar{M}^2_{r})\right],
\cr
&\approx&
-\frac{1}{(2\pi)^4}\frac{M_kM_r}{2F^2_\phi}
\left(\frac{\rlap{/}{k}+\bar{M}_f}{k^2-\bar{M}^2_f}\right)\gamma_5
\left(\rlap{/}{r}+\bar{M}_{f'}\right)\gamma_5
\left(\frac{\rlap{/}{k}+\bar{M}_f}{k^2-\bar{M}^2_f}\right)
\left[2\pi\delta(r^2-\bar{M}^2_{f'})\right],
\end{eqnarray}
where the momentum-dependent effective quark mass is written via Eq.~(\ref{eq:MDM}):
\begin{equation}
\label{eq:MDQM}
M_\ell=M_0\left[\frac{2\Lambda^2}{2\Lambda^2-\ell^2+i\epsilon} \right]^{2}.
\end{equation}
Note that the $\mu$ dependence of $\Delta(k,p,\mu)$ is due to that fact we need regularize $M_\ell$ by introducing the model renormalization 
scalar $\Lambda$ in Eq. (\ref{eq:MDM}).
In Eq.~(\ref{eq:DELTA}) we adopt the following notations: $\bar{M}_\ell\equiv m_f+M_\ell$ and $\bar{M}_f\equiv m_f+M_0$. Note that $f$ and $f'$ indicate the flavors for the initial $(q)$ and final $(q')$ quarks, respectively. Eq.~(\ref{eq:DELTA}) stands for the $\pi^+$ elementary fragmentation, i.e. $u\to\pi^+\,d$. As for the fragmentation of a neutral pion, one needs multiply {the isospin factor $1/2$ to Eq.~(\ref{eq:DELTA}) at the elementary-fragmentation level}. From the first line to the second line in Eq.~(\ref{eq:DELTA}), we approximate some of the momentum-dependent effective quark masses into a constant constituent-quark mass $M_\ell\to M_0$, since the momentum dependencies from those mass terms play only minor roles numerically. Moreover, this approximation simplifies the numerical calculations to a great extent. Performing the trace over the Lorentz indices for the fragmentation function in Eq.~(\ref{eq:FRAG}):
\begin{equation}
\label{eq:TRACE}
\mathrm{Tr}\left[\left(\rlap{/}{k}+\bar{M}_f\right)\gamma_5
\left(\rlap{/}{k}-\rlap{/}{p}+\bar{M}_{f'}\right)\gamma_5
\left(\rlap{/}{k}+\bar{M}_f\right)\gamma_{\mu} \right]
=-4\left[p_\mu(k^2-\bar{M}^2_f)+k_\mu(\bar{M}^2_f-2\bar{M}_f\bar{M}_{f'}
+k^2-2k\cdot p) \right],
\end{equation}
we reach a concise expression for the elementary fragmentation function $u\to\pi^+ d$ from NLChQM:
\begin{equation}
\label{eq:FRAGDEF}
d^{\pi^+}_{u}(z,\bm{k}^2_T,\mu)=\frac{1}{8\pi^3z(1-z)}\frac{M_kM_{r}}{2F^2_\pi}
\left[\frac{z(k^2-\bar{M}^2_f)+(k^2+\bar{M}^2_f-2\bar{M}_f\bar{M}_{f'}-2k\cdot p) }{(k^2-\bar{M}^2_f)^2}\right],
\end{equation}
where the relevant four momenta squared in the present theoretical calculations are defined in the light-cone coordinate as follows:
\begin{equation}
\label{eq:FOUR}
k^2=k_+k_--\bm{k}^2_T,\,\,\,\,r^2=(k-p)^2\approx k_+k_--\bm{k}^2_T+m^2_\phi-(k_+p_-+k_-p_+),
\nonumber
\end{equation}
where the transverse PS-meson momentum on the light cone, i.e. $\bm{p}_T=0$. {Note that the value of $M_0$ can be fixed self-consistently within the instanton model~\cite{Diakonov:1985eg,Shuryak:1981ff,Diakonov:1983hh,Schafer:1996wv,Musakhanov:1998wp,Musakhanov:2002vu,Diakonov:2002fq,Nam:2007gf,Nam:2010pt} with the phenomenological (anti)instanton parameters $\bar{\rho}\approx1/3$ fm and $\bar{R}\approx1$ fm, resulting in $M_0\approx350$ MeV by the following self-consistent equation, defined in Euclidean ($E$) space:
\begin{equation}
\label{eq:SELF}
\frac{1}{\bar{R}^4}\approx4N_c\int_E\frac{d^4q}{(2\pi)^4}\frac{M(q^2)}{q^2+M(q^2)},
\,\,\,\,M(q^2)=M_0\left[\frac{2}{2+\bar{\rho}^2q^2} \right]^2.
\end{equation}
The pion mass is chosen to be $m_\pi=140$ MeV throughout the present work}. Collecting all the ingredients above, one is led to a final expression for the elementary fragmentation function:
\begin{equation}
\label{eq:DDDDD}
d^{\pi^+}_{u}(z,\bm{k}^2_T,\mu)=\frac{1}{8\pi^3}\frac{M_kM_{r}}{2F^2_\pi}
\frac{z\left[z^2\bm{k}^2_T+[(z-1)\bar{M}_f+\bar{M}_{f'}]^2\right]}
{[z^2\bm{k}^2_T+z(z-1)\bar{M}^2_f+z\bar{M}^2_{f'}+(1-z)m^2_\pi]^2},
\end{equation}
where $M_k$ and $M_{r}$ are the momentum-dependent quark mass manifesting the nonlocal quark-PS meson interactions, read:
\begin{equation}
\label{eq:MASS}
M_k=\frac{M_0[2\Lambda^2z(1-z)]^2}
{[z^2\bm{k}^2_T+z(z-1)(2\Lambda^2-\delta^2)+z\bar{M}^2_{f'}+(1-z)m^2_\pi]^2},
\,\,\,\,
M_{r}=\frac{M_0(2\Lambda^2)^2}{(2\Lambda^2-\bar{M}^2_{f'})^2}.
\end{equation}
As for $M_k$ in Eq.~(\ref{eq:MASS}), we have introduced a free and finite-valued parameter $\delta$ in the denominator to avoid the unphysical singularities, which appear in the vicinity of $(z,\bm{k}_T)=0$, due the present parametrization of the effective quark mass as in Eq.~(\ref{eq:MDM}). We determine $\delta$ to satisfy a natural conditions for all the possible values of $(z,\bm{k}_T)$:
\begin{equation}
\label{eq:CON}
|z^2\bm{k}^2_T+z(z-1)(n\Lambda^2-\delta^2)+z\bar{M}^2_{f'}+(1-z)m^2_\pi|>0.
\end{equation}
It is easy to see that one of the trivial solutions for Eq.~(\ref{eq:CON}) can be $\delta^2=n\Lambda^2-\bar{M}^2_f$, considering the quark propagator in Eq.~(\ref{eq:DDDDD}). Although there can be other solutions, we will use this value for $\delta$ for the fragmentation function as a trial hereafter. Physically, those unphysical singularities can be understood as that a hypothetic particle, whose mass corresponds to $\sqrt{2}\Lambda$, becomes its on-mass shell for a certain combination of $(z,\bm{k}_T)$. Thus, we employed $\delta$ to exclude that unphysical situation. We also verified that the change of $\delta$ does not make considerable effects on the numerical results, as far as Eq.~(\ref{eq:CON}) is fulfilled.

If we replace all the momentum-dependent masses into a constant one $M$ in Eqs.~(\ref{eq:DDDDD}) and (\ref{eq:MASS}) in the chiral limit, and change the quark-PS meson coupling into a constant one appropriately in Eq.~(\ref{eq:DDDDD}), we obtain the expression for the elementary fragmentation function from the quark-meson coupling model~\cite{Amrath:2005gv} :
\begin{equation}
\label{eq:DDDDD1}
d^{\pi^+}_{u}(z,\bm{k}^2_T,\mu)=\frac{1}{8\pi^3}g^2_{\phi qq}
\frac{z(z^2\bm{k}^2_T+z^2M^2)}{[z^2\bm{k}^2_T+z^2M^2+(1-z)m^2_\pi]^2}.
\end{equation}
The above equation is also equivalent with that from the NJL model calculations in principle~\cite{Bentz:1999gx,Ito:2009zc,Matevosyan:2011ey}. Note that the on-shell value of $M^2_0/(2F^2_\pi)$ becomes about $7$, which is quite similar to $g^2_{qq\pi}\approx9$, used in Refs.~\cite{Amrath:2005gv,Bentz:1999gx,Ito:2009zc,Matevosyan:2011ey}. At the renormalization scale in our model, the elementary fragmentation function is assumed to be able to be evaluated further by integrating Eq.~(\ref{eq:DDDDD}) over $k_T$ as:
\begin{equation}
\label{eq:FRAGINT}
d^{\pi^+}_{u}(z,\mu)=2\pi z^2\int^\infty_0 d^{\pi^+}_{u}(z,\bm{k}^2_T,\mu)\,\bm{k}_T\,d\bm{k}_T,
\end{equation}
where the factor $z^2$ in the right-hand-side is again comes from the integration over $\bm{p}_\perp=-\bm{k}_T/z$. Actually the connection between $d^{\pi^+}_{u}(z,\mu)$ and $d^{\pi^+}_{u}(z,\bm{k}^2_T,\mu)$ would be far more complicated in principle~\cite{Aybat:2011zv}.} We will present the numerical results for Eq.~(\ref{eq:FRAGINT}) in the next Section.

Using the DLY relation in Eq.~(\ref{eq:DLY}), one can derive the quark-distribution function for $\pi^+$, i.e. $f^{\pi^+}_{u}$ as follows:
\begin{equation}
\label{eq:PDF0}
f^{\pi^+}_{u}(x,\bm{k}^2_T,\mu)=\frac{3}{4\pi^3}
\frac{\mathcal{M}_k\mathcal{M}_{r}}{2F^2_\pi}
\frac{\bm{k}^2_T+[(x-1)\bar{M}_f-x\bar{M}_{f'}]^2}
{[\bm{k}^2_T+(1-x)\bar{M}^2_f+x\bar{M}^2_{f'}+x(x-1)m^2_\pi]^2},
\end{equation}
where we have defined the effective quark masses for the quark distribution function by
\begin{equation}
\label{eq:EFFFFF}
\mathcal{M}_k=\frac{4M_0\Lambda^4(1-x)^2}
{[\bm{k}^2_T+(1-x)(2\Lambda^2)+x\bar{M}^2_{f'}+x(x-1)m^2_\pi]^2},\,\,\,\,
\mathcal{M}_r=\frac{4M_0\Lambda^4}{(2\Lambda^2-\bar{M}^2_{f'})^2}.
\end{equation}
Note that $\mathcal{M}_k$ in Eq.~(\ref{eq:EFFFFF}) does not suffer from the unphysical singularities unlike that in Eq.~(\ref{eq:MASS}), according to the different kinematic situations between those functions. Again, if we take the replacements as $\bar{M}_{f,f'}\to M$ and $\mathcal{M}_k\mathcal{M}_r/(2F^2_\pi)\to g^2_{\phi qq}$ as done for the fragmentation function, one is led to
\begin{equation}
\label{eq:PDF}
f^{\pi^+}_{u}(x,\bm{k}^2_T,\mu)=\frac{3}{4\pi^3}g^2_{\phi qq}
\frac{\bm{k}^2_T+M^2}{[\bm{k}^2_T+M^2+x(x-1)m^2_\pi]^2},
\end{equation}
which is equivalent to Eq.~(14) in Ref.~\cite{Ito:2009zc} in the chiral limit. Similarly, the integration over $k_T$ can be performed as follows, resulting in a function of $x$ at a certain renormalization scale $\mu\approx\Lambda$:
\begin{equation}
\label{eq:PDF2}
f^{\pi^+}_{u}(x,\mu)=2\pi\int^\infty_0f^{\pi^+}_{u}(x,\bm{k}^2_T,\mu)\,\bm{k}_T\,d\bm{k}_T.
\end{equation}
We will use the notation $\pi$ instead of $\pi^+$ from now on for convenience. {Moreover, the minus-type quark-distribution function, which corresponds to the valance-quark distribution, satisfies the following normalization condition:
\begin{equation}
\label{eq:PDFNORM}
\int^1_0 dx\,\left[f^{\pi^+}_u(x,\mu)-f^{\pi^+}_{\bar{u}}(x,\mu) \right]=n_q,
\end{equation}
where $n_q$ in the right-hand-side indicates the valance-quark number and becomes unity.}
\section{Numerical results}
Here we present our numerical results for the fragmentation functions and the parton distribution functions of $\pi$ with relevant discussions. We will consider only the favored fragmentation process $u \to\pi  d$. As in the instanton model the value of $\Lambda$ is proportional to $1/\bar{\rho}$ giving the renormalization scale $\Lambda\approx600$ MeV as mentioned in the previous section. {As understood by Eq.~(\ref{eq:MDQM}), the effective quark mass plays the role of the UV regulator by construction, resulting in that relevant integrals in Eqs.~(\ref{eq:FRAGINT}) and (\ref{eq:PDF2}) are not divergent. This feature is quite different from Refs.~\cite{Amrath:2005gv,Bentz:1999gx,Ito:2009zc,Matevosyan:2011ey}, in which various cutoff scheme were employed. Moreover, we assume that there is no perturbative gluon contributions for those function in hand at the initial scale $\Lambda$.

First, we present the numerical results for the elementary fragmentation functions, multipiled by $z$, $zd^\pi_u$ as functions of $z$ in the left panel of Figure~\ref{FIG1} for the present (solid), NJL (dot), PS (dash), and PV (dot-dash) results. As for the NJL result, the renormalization scale is chosen to be $0.18\,\mathrm{GeV}^2$ as usual~\cite{Amrath:2005gv,Bentz:1999gx,Ito:2009zc,Matevosyan:2011ey}, whereas it becomes $1\,\mathrm{GeV}^2$ for the PS and PV schemes of the quark-meson coupling model~\cite{Amrath:2005gv}. {It is worth mentioning that the initial $Q^2$ values for each curve are chosen to be the same with each model renormalization scales. For instance, we set $Q^2=(600\,\mathrm{MeV})^2$ for the present case. Although, in order for an appropriate comparison between the model results, one may need proper $Q^2$ evolutions to a single $Q^2$ value, we do not perform those evolutions here. Hence, the curves in Figure~\ref{FIG1} should be taken into account as a very qualitative comparison at a generic hadron scale $(0.4\sim1.0)$ GeV.} From the left panel of Figure~\ref{FIG1}, we observe that our curves are in general much larger than the other model results. Moreover, while the peaks of the curves locate at $z\sim0.5$ for the present theoretical framework, the maximum values of NJL and PS scheme are around $z\sim0.7$. On the other hand, the maximum value occurs in PV scheme at the location where is close to ours. The behaviors of the NJL and PS-scheme curves are similar because they are both local models without derivatives.

Note that these elementary fragmentation functions computed here do not satisfy the momentum sum rule Eq.~(\ref{eq:SUM}).
Since we have considered the elementary process $q\to\pi\,q'$, however, as discussed in Refs.~\cite{Ito:2009zc,Matevosyan:2011ey}, the pion-cloud effects will enhance the elementary fragmentation functions, and those fragmentation functions with the pion-cloud effects were identified as the {\it renormalized} fragmentation function. The pion-cloud effects can be estimated by the value of the first moment of the fragmentation functions:
\begin{equation}
\label{eq:POP}
1-Z_{Q}=\sum_\pi\int^1_0dz\,d^\pi_u(z)\equiv \mathcal{N}_{\pi/u}.
\end{equation}
where the summation runs over the all isospin states of the PS meson. For instance, in Refs.~\cite{Ito:2009zc,Matevosyan:2011ey}, the value of $\mathcal{N}_{\pi/u}$ was estimated to be $(0.1\sim0.2)$. As understood, this smallness of $\mathcal{N}_{\pi/u}$ results in the breakdown of the momentum sum rule. The renormalized fragmentation function then can be defined as~\cite{Ito:2009zc}:
\begin{equation}
\label{eq:RENORM}
D^\pi_u(z)=\frac{ d^\pi_u(z)}{\mathcal{N}_{\pi/u}}.
\end{equation}
Here, we note that the summation runs over all the possible fragmented hadron states from the $u$ quark in principle. However, since we are interested only in the pion case, we restrict ourselves to the SU(2) light-flavor mesons, i.e. no kaons.  From the numerical computations, we obtain
\begin{equation}
\label{eq:RENOMV}
\mathcal{N}^{\mathrm{present}}_{\pi/u}=0.3685,\,\,\,\,
\mathcal{N}^{\mathrm{NJL}}_{\pi/u}=0.0257,\,\,\,\,
\mathcal{N}^{\mathrm{PS}}_{\pi/u}=0.0843,\,\,\,\,
\mathcal{N}^{\mathrm{PV}}_{\pi/u}=0.1684.
\end{equation}
From the values in Eq.~(\ref{eq:RENOMV}), we find that present result gives much larger strength compared wit those in the models with local interactions. This observation tells us that the probability for the initial quark to be fragmented into the hadrons, especially pion, turns out to be much higher due to the nonlocal interactions. Physically, this enhancement of $\mathcal{N}_{\pi/q}$ can be understood in the following way: The nonlocal interactions are originally generated from the intricate quark-(anti)instanton interactions in the instanton-vacuum picture~\cite{Diakonov:1985eg,Shuryak:1981ff,Diakonov:1983hh,Schafer:1996wv,Diakonov:2002fq}. This momentum-dependent quark mass can be understood as a dressed-quark mass as in usual Dyson-Schwinger methods~\cite{Nguyen:2011jy} where the quarks are dressed by the pion cloud. In this sense, the nonlocal interaction represented by the momentum-dependent effective quark mass practically contains considerable contributions from the pion cloud. It leads to the higher probabilities for the initial quark to be fragmented into the pion, comparing to usual local models with a constant quark mass and coupling. The numerical results for the renormalized fragmentation function multiplied by $z$, $ zD^{\pi}_u(z)$ is given in the right panel of Figure~\ref{FIG1}.  Interestingly, the curves from the present nonlocal model calculations look like an average of the PS- (or NJL) and PV-scheme results. This can be understood qualitatively by expanding the momentum-dependent quark masses of $\sqrt{M_{k}M_{r}}\sim M_k$ in Eq.~(\ref{eq:MDQM}):
\begin{equation}
\label{eq:EXPMDQM}
M_k\approx M_0-M_0\frac{k^2}{\Lambda^2}+\cdots.
\end{equation}
The first term of Eq.~(\ref{eq:EXPMDQM}) relates to the PS scheme (or NJL) without derivatives, while the second term to the PV one with derivatives, qualitatively. In this way, the present result can be seen as sort of a mixture of the two coupling models.
\begin{figure}[t]
\begin{tabular}{cc}
\includegraphics[width=8.5cm]{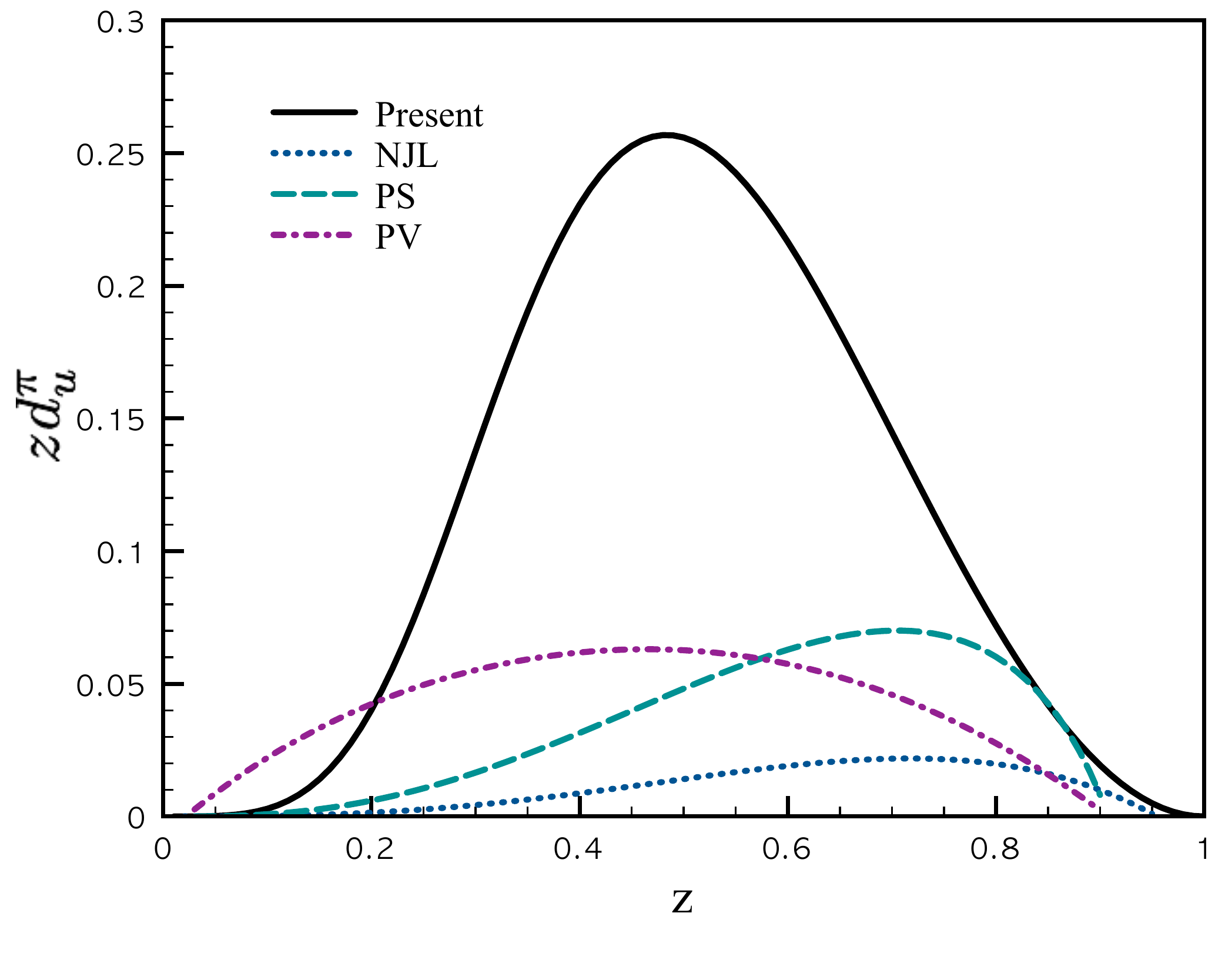}
\includegraphics[width=8.5cm]{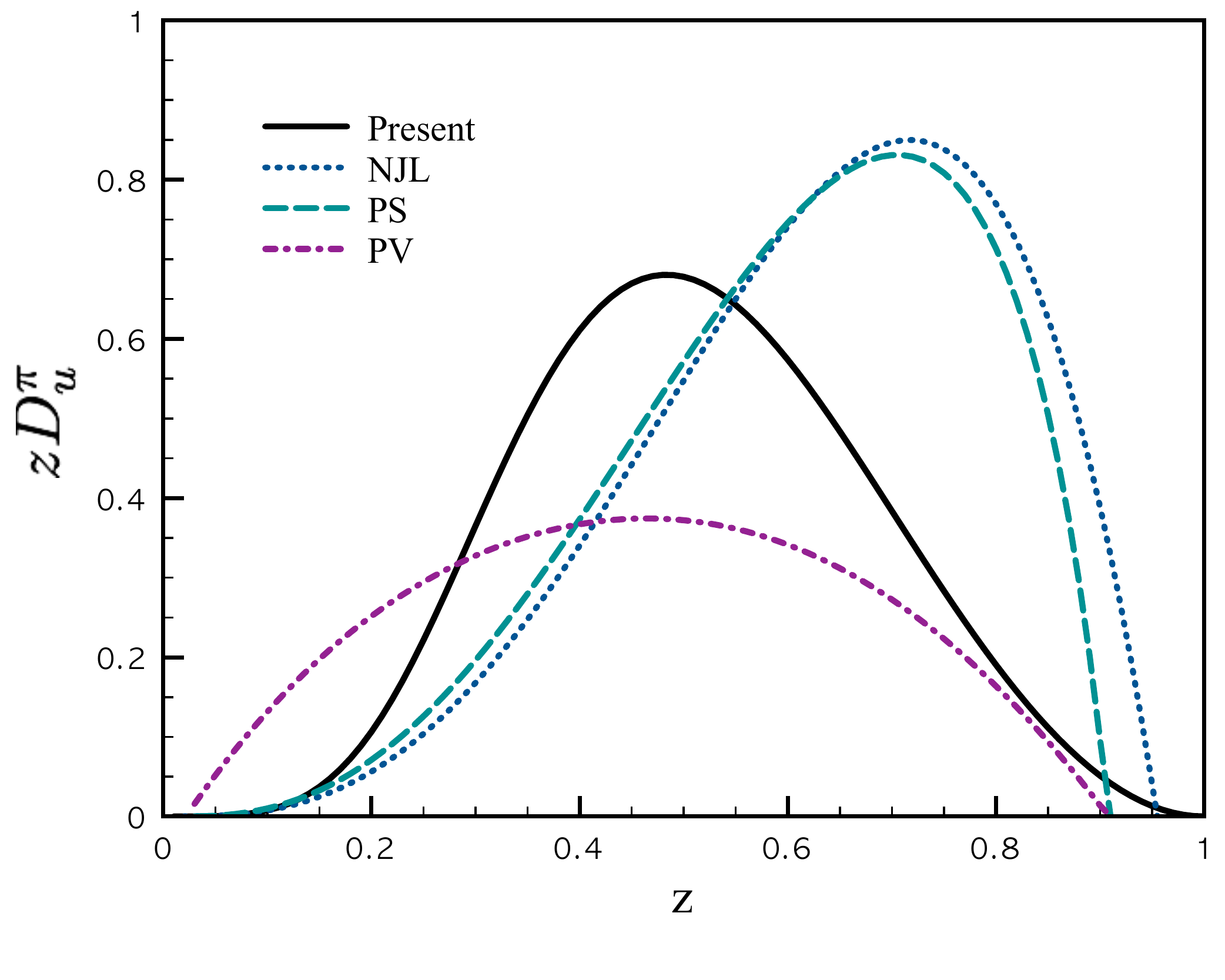}
\end{tabular}
\caption{(Color online) In the left panel, we show $zd^\pi_u(z)$ at low renormalization scale for the present result ($Q^2=0.36\,\mathrm{GeV}^2$, solid), NJL~\cite{Ito:2009zc,Matevosyan:2011ey} ($Q^2=0.18\,\mathrm{GeV}^2$, dot), PS scheme~\cite{Bacchetta:2002tk,Amrath:2005gv} ($Q^2=1\,\mathrm{GeV}^2$, dash), and PV scheme~\cite{Bacchetta:2002tk,Amrath:2005gv} ($Q^2=1\,\mathrm{GeV}^2$, dot-dash). $zD^\pi_u(z)$ using Eq.~(\ref{eq:RENORM}) are also shown in the right panel in the same manner with the left panel.}
\label{FIG1}
\begin{tabular}{cc}
\includegraphics[width=8.5cm]{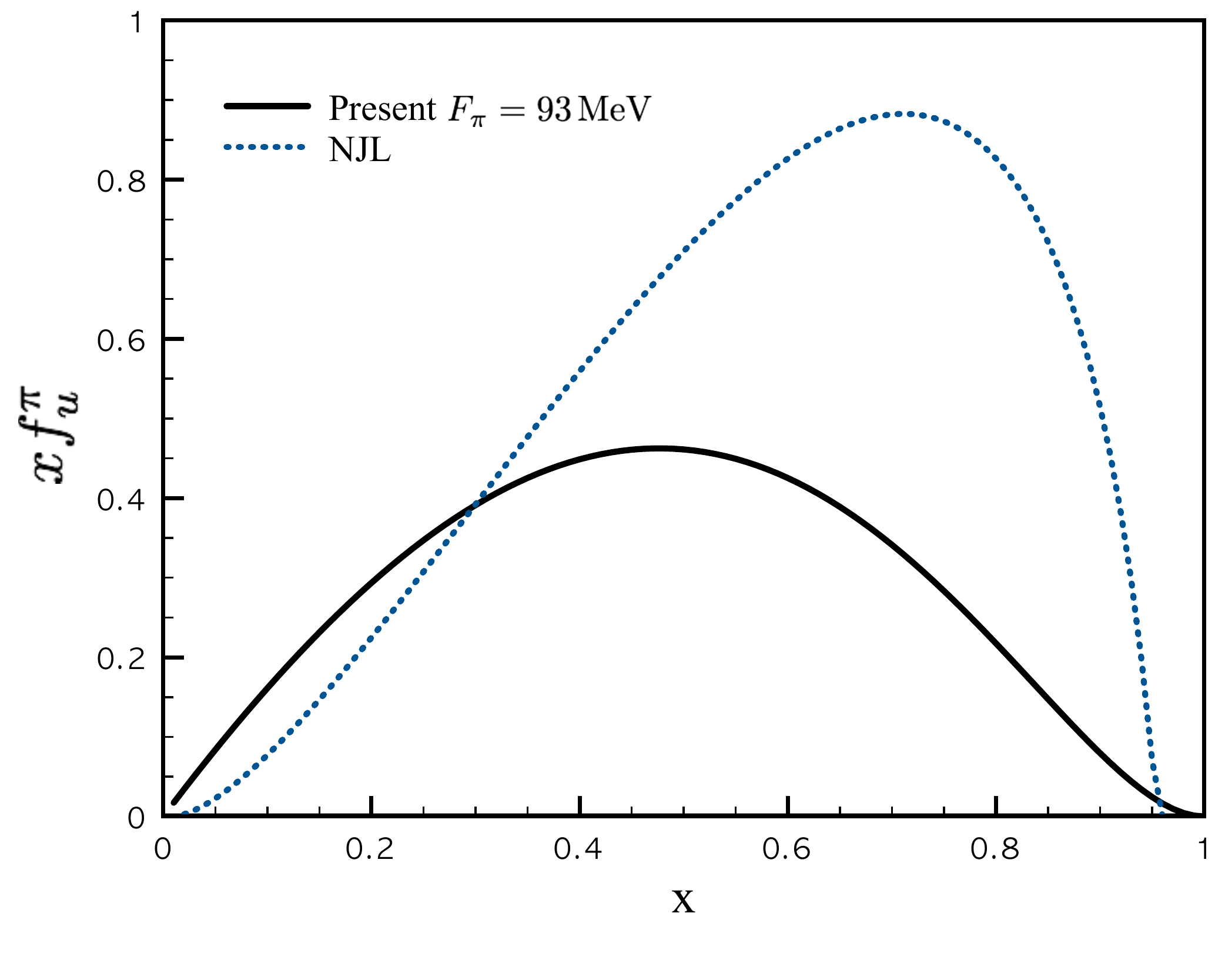}
\includegraphics[width=8.5cm]{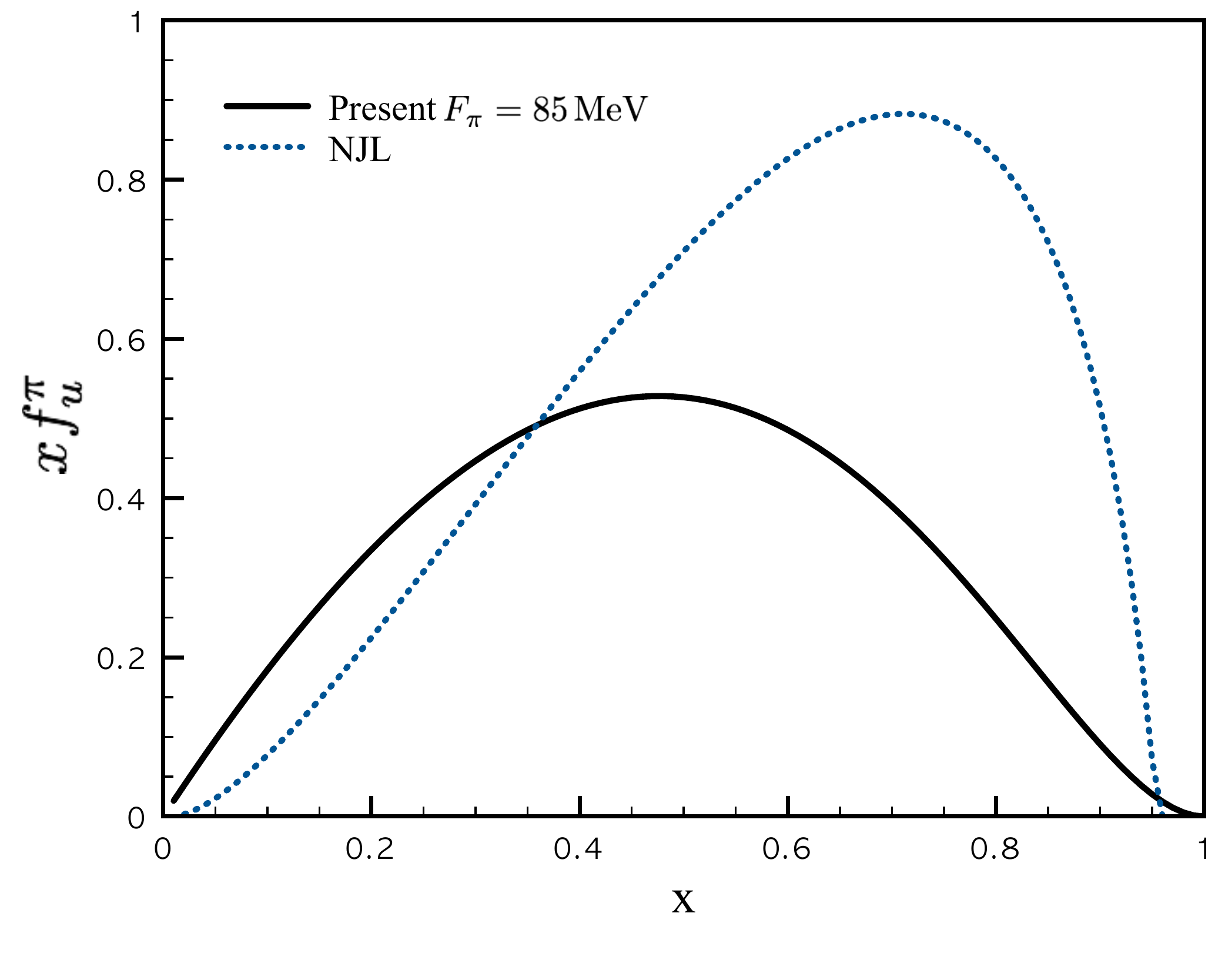}
\end{tabular}
\caption{(Color online) $xf^\pi_u(x)$ from the present model (solid), and NJL~\cite{Ito:2009zc,Matevosyan:2011ey} (dot) for $F_\pi\approx93$ MeV (left) and $85$ MeV (right) at low $Q^2$ values: $Q^2=0.36\,\mathrm{GeV}^2$ for the present models and $Q^2=0.18\,\mathrm{GeV}^2$ for the NJL.}
\label{FIG2}
\end{figure}

The numerical results for $xf^\pi_u$ are given in Figure~\ref{FIG2} for the present result (solid) and NJL (dot) cases as functions of the momentum fraction $x$. All the calculations are carried out at the same renormalization scale $600$ MeV as for the fragmentation functions. {If we integrate the quark-distribution function over $x$ as in Eq.~(\ref{eq:PDFNORM}) with the empirical pion-decay constant $F_\pi\approx93$ MeV for Eq.~(\ref{eq:PDF0}), shown in the left panel of Figure~\ref{FIG2}, we have $n_q=0.88$. These values deviate the normalization condition of Eq.~(\ref{eq:PDFNORM}) by $(10\sim20)\%$. The reason for this discrepancy can be understood as follows: Within NLChQM (and also the instanton model), the value of $F_\pi$ turns out to be about $20\%$ smaller than its empirical value, if the mass-derivative terms, i.e. $\partial M_k/\partial |k|$, are ignored~\cite{Diakonov:1985eg,Shuryak:1981ff,Diakonov:1983hh,Schafer:1996wv,Musakhanov:1998wp,Musakhanov:2002vu,Diakonov:2002fq,Nam:2007gf,Nam:2010pt}. Note that these terms are originated by considering the (axial) vector current conservation for the relevant hadronic matrix elements~\cite{Diakonov:1985eg,Bowler:1994ir,Nam:2008xx}. Practically, they are called as {\it nonlocal contributions}. In deriving $d^{\pi}_u$ in Section II, however, we did not include them and derived $f^\pi_u$ by employing the DLY relation. Consequently our values of $f^\pi_u$ violate the normalization condition Eq.~(\ref{eq:PDFNORM}). Thus, if we choose a smaller value for $F_\pi$, considering the absence of the nonlocal contributions, the normalization condition can be satisfied with $F_\pi\approx85$ MeV. In other words, the effects from the nonlocal contributions are compensated by the smaller $F_\pi$ value. The inclusion of the nonlocal contributions is under progress~\cite{NEXT}. The numerical results which satisfy Eq.~(\ref{eq:PDFNORM}) are given in the right panel of Figure~\ref{FIG2}. As shown there, the overall strength of the curves are enhanced by about $10\%$, whereas the curves shapes remain almost the same. From a phenomenological point of view, we will use the numerical results for $f^\pi_u$, satisfying the normalization, given in the right panel of Figure~\ref{FIG2}, hereafter. It also turns out that the curve from the present calculation is very symmetric and peak at $x\approx0.5$. The NJL result is tilted to the the region near $x=1$. Thus, we conclude that the nonlocal interactions between the quark and PS meson have large impact on the shape of the curves of $xf^\pi_u$ in comparison to those from the local-interaction model calculations, as already observed in the case of the fragmentation functions in Figure~\ref{FIG1}.

According to Ref.~\cite{Sutton:1991ay}, $xf^\pi_u$ can be parameterized by the following forms:
\begin{eqnarray}
\label{eq:QDFPARASUT}
xf^{\pi}_{u}(x)&=&xf^{\pi}_{\bar{d}}(x)=A^{\pi}_vx^{\alpha_\pi}(1-x)^{\beta_\pi}
\,:\,\mathrm{quark\,\,contribution},
\cr
xf^{\pi}_{\bar{u}}(x)&=&xf^{\pi}_{d}(x)=\frac{1}{3}A^{\pi}_s(1-x)^{\eta_\pi}
\,:\,\mathrm{antiquark\,\,contribution},
\end{eqnarray}
where we have assumed that SU(2) flavor symmetry for the valance and sea quarks, respectively. {Using Eq.~(\ref{eq:QDFPARASUT}), $xf^{\pi}_{u}(x)$ in Figure~\ref{FIG2} can be parameterized at $Q^2=0.36\,\mathrm{GeV}^2$ and the fitted values for $A^\pi_v$, $\alpha_\pi$, and $\beta_\pi$ are listed in Table~\ref{TABLE1}, satisfying the normalization condition in Eq.~(\ref{eq:PDFNORM}).} These parameterized distribution functions will be used for the inputs for the high-$Q^2$ evolution. To compare with the empirical data obtained in Ref.~\cite{Hirai:2007cx}, we need to perform the high-$Q^2$ evolution of the present results. For this purpose, we will make use of the numerical DGLAP evolution code {\it QCDNUM17} by Botje~\cite{Botje:2010ay,DGLAP}. The quark distribution functions computed and parameterized previously becomes the inputs for the evolution. For the QCDNUM evolution, one needs three valance quark $x(q-\bar{q})$ and anti-quark $x\bar{q}$ distributions.
Note that we assume that the gluon functions is zero at the initial scale.
\begin{table}[h]
\begin{tabular}{c||c|c|c}
$Q^2=0.36\,\mathrm{GeV}^2$&$A^\pi_v$&$\alpha_\pi$&$\beta_\pi$\\
\hline
Present&$\,\,\,\,3.3818\,\,\,\,$&\,\,\,\,$1.2163$\,\,\,\,&\,\,\,\,$1.4280$\,\,\,\,\\
NJL&$5.2497$&$1.9131$&$0.9208$\\
\end{tabular}
\caption{Fitted values for the quark distribution function, $A^\pi_v$, $\alpha_\pi$, and $\beta_\pi$ in Eq.~(\ref{eq:QDFPARASUT}) at $Q^2=0.36\,\mathrm{GeV}^2$.}
\label{TABLE1}
\end{table}

First, in Figure~\ref{FIG3}, we show the numerical results for the renormalized $zD^\pi_u$ evolved from $Q^2=0.36\,\mathrm{GeV}^2$ (solid) to $Q^2=1\,\mathrm{GeV}^2$ at LO (dot) evolution. {The empirical fragmentation function is parametrized as follows~\cite{Hirai:2007cx}}:
\begin{equation}
\label{eq:HIRAI}
D^\pi_u(z)=N^\pi_u\,z^{\alpha^\pi_u}\,
(1-z)^{\beta^\pi_u},
\,\,\,\,
N^\pi_u=\frac{M^\pi_u}{\mathrm{B}
(\alpha^\pi_u+2,\,\beta^\pi_u+1)},
\end{equation}
where $\mathrm{B}(x,y)$ in the denominator stands for the beta function with arguments $x$ and $y$. The LO and NLO values for $\alpha$ and $\beta$, obtained at $Q^2=1\,\mathrm{GeV}^2$ are given in Table~\ref{TABLE2} following Ref.~\cite{Hirai:2007cx}.
\begin{table}[h]
\begin{tabular}{c||c|c|c||c|c|c}
&$M^\pi_u$ (LO)&$\alpha^\pi_u$ (LO)&$\beta^\pi_u$ (LO)
&$M^\pi_u$ (NLO)&$\alpha^\pi_u$ (NLO)&$\beta^\pi_u$ (NLO)\\
\hline
$D^{\pi}_{u}(z)$&$0.546\pm0.085$&$-1.100\pm0.183$&$1.282\pm0.140$
&$0.401\pm0.052$&$-0.963\pm0.177$&$1.370\pm0.144$\\
\end{tabular}
\caption{Parameters for the empirical pion fragmentation functions in Eq.~(\ref{eq:HIRAI}), evaluated at $Q^2=1\,\mathrm{GeV}^2$ for LO and NLO evolutions, in Ref.~\cite{Hirai:2007cx}.}
\label{TABLE2}
\end{table}

The empirical results for $zD^\pi_u$ at $Q^2=1\,\mathrm{GeV}^2$ at LO and NLO are drawn in Figure~\ref{FIG3} in the dash and long-dash lines, respectively with the inputs including Eq.~(\ref{eq:HIRAI}) and the values listed in Table~\ref{TABLE2}. {The errors for the empirical fragmentation functions are also depicted.} As for the present result (left), the LO and NLO empirical curves are qualitatively matching with the theoretical calculation (dot) for the region $z\gtrsim0.5$ whereas they deviate considerably in the region $z\lesssim0.5$. As discussed in Refs.~\cite{Ito:2009zc,Matevosyan:2011ey}, the fragmentation functions will be enhanced by taking into account the quark-jet contributions in the region $z\lesssim0.5$. Hence, although we have not taken those contributions into account here, the discrepancy between the theoretical and empirical curves might be cured by including the jet contributions which are now under progress ~\cite{NEXT}.
\begin{figure}[t]
\includegraphics[width=8.5cm]{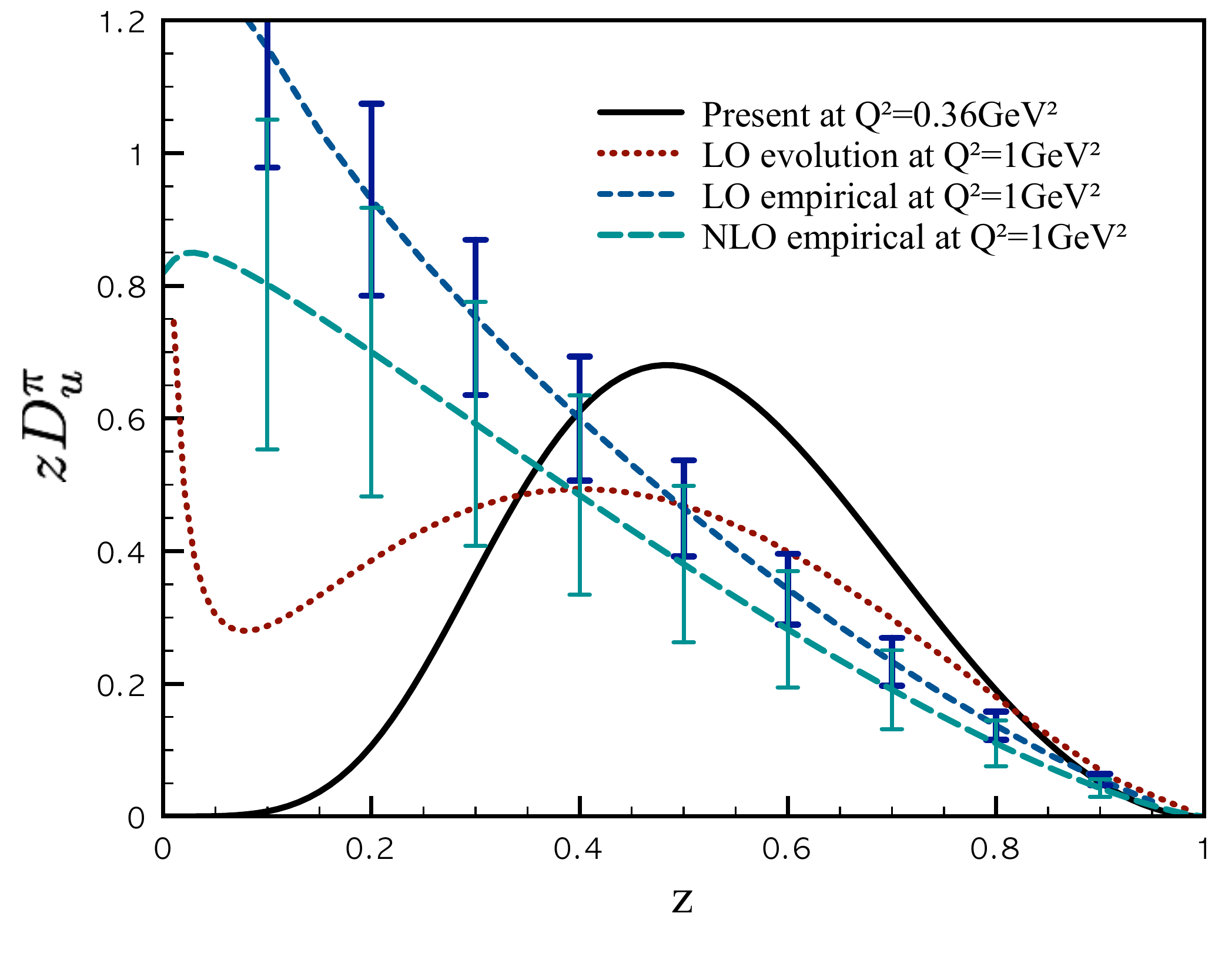}
\caption{(Color online) Renormalized elementary fragmentation functions, multiplied by $z$, $zD^\pi_u(z)$ at $Q^2=0.36\,\mathrm{GeV}^2$ (solid) and they are evolved to $Q^2=1\,\mathrm{GeV}^2$ at LO (dot). The empirical data are taken from Ref.~\cite{Hirai:2007cx} for the LO (dash) and NLO (long-dash) analyses at $Q^2=1\,\mathrm{GeV}^2$.}
\label{FIG3}
\end{figure}

We demonstrate our results evolved from $Q^2=0.36\,\mathrm{GeV}^2$ (solid) to $4\,\mathrm{GeV}^2$ at LO (dot) and NLO (dash) in Figure~\ref{FIG4} for the minus-type (left column) and plus-type (right column). The empirical curves are computed using Eq.~(\ref{eq:QDFPARASUT}) and input values fitted from the NA10 data~\cite{Sutton:1991ay} in Table~\ref{TABLE3}. {Note that we did not present the errors for the empirical data in Figure~\ref{FIG4}, since the errors given in Ref.~\cite{Sutton:1991ay} provide considerably small effects.} It is clearly shown that the present results are in agreement with the empirical one for the minus-type distribution, which is nothing but the valance quark distribution function. We also observe that the present results for the plus-type distribution functions underestimate in the region $x\lesssim0.2$ .
\begin{table}[h]
\begin{tabular}{c|c|c|c}
\,\,\,\,\,$\alpha_\phi$\,\,\,\,\,
&\,\,\,\,\,$\beta_\phi$\,\,\,\,\,
&\,\,\,\,\,$A^\phi_s$\,\,\,\,\,
&\,\,\,\,\,$\eta_\phi$\,\,\,\,\,\\
\hline
$0.64\pm0.03$&$1.08\pm0.22$&$0.9\pm0.3$&$5.0$\\
\end{tabular}
\caption{Input values for the empirical quark distribution function for Eq.~(\ref{eq:QDFPARASUT}), given in Ref.~\cite{Sutton:1991ay}.}
\label{TABLE3}
\end{table}

\begin{figure}[t]
\begin{tabular}{cc}
\includegraphics[width=8.5cm]{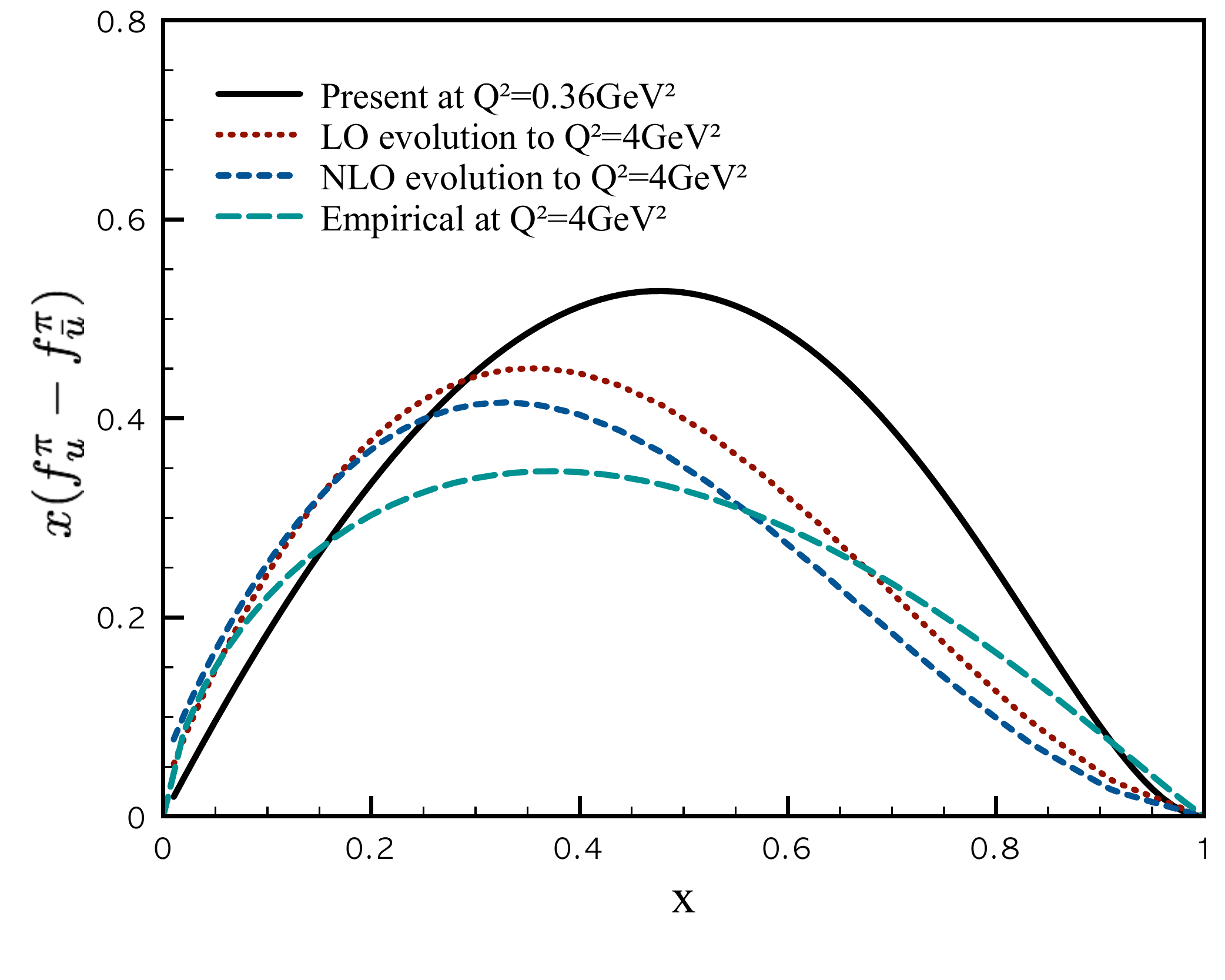}
\includegraphics[width=8.5cm]{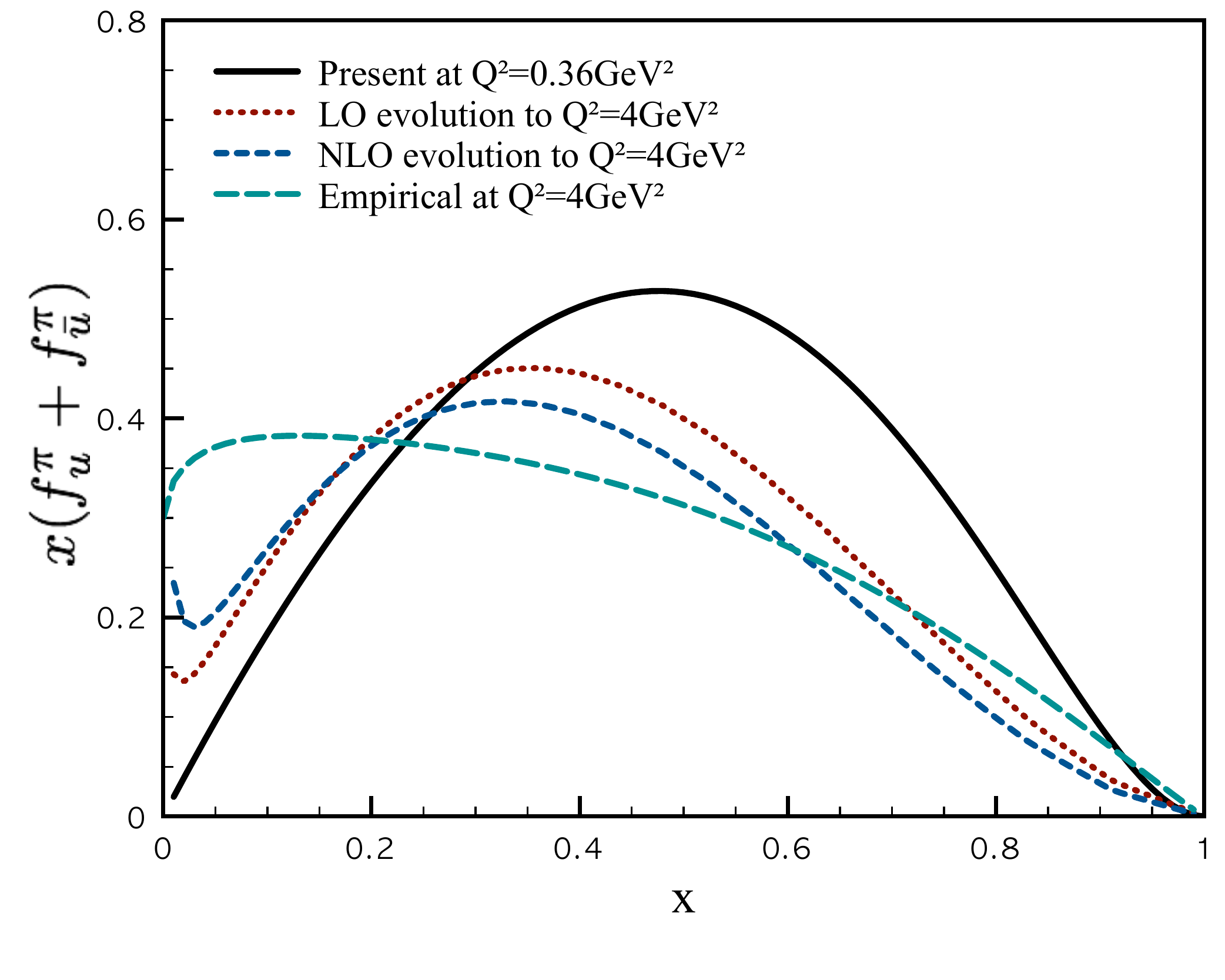}
\end{tabular}
\caption{(Color online) Minus-type distribution functions $x(f^\pi_u-f^\pi_{\bar{u}})$ (left) and plus-type $x(f^\pi_u+f^\pi_{\bar{u}})$ (right) at different $Q^2$ values, i.e. $Q^2=0.36\,\mathrm{GeV}^2$ and $4\,\mathrm{GeV}^2$. The $Q^2$ evolution is performed at LO and NLO. The empirical data are taken from Ref.~\cite{Sutton:1991ay}.}
\label{FIG4}
\end{figure}

After the evolution, we plot the valance quark distribution function at $Q^2=27\,\mathrm{GeV}^2$ in Figure~\ref{FIG5}. The data points are taken from Ref.~\cite{Conway:1989fs}. The present result reproduces the data qualitatively well as expected from the left panel of Figure~\ref{FIG4}.
\begin{figure}[t]
\includegraphics[width=8.5cm]{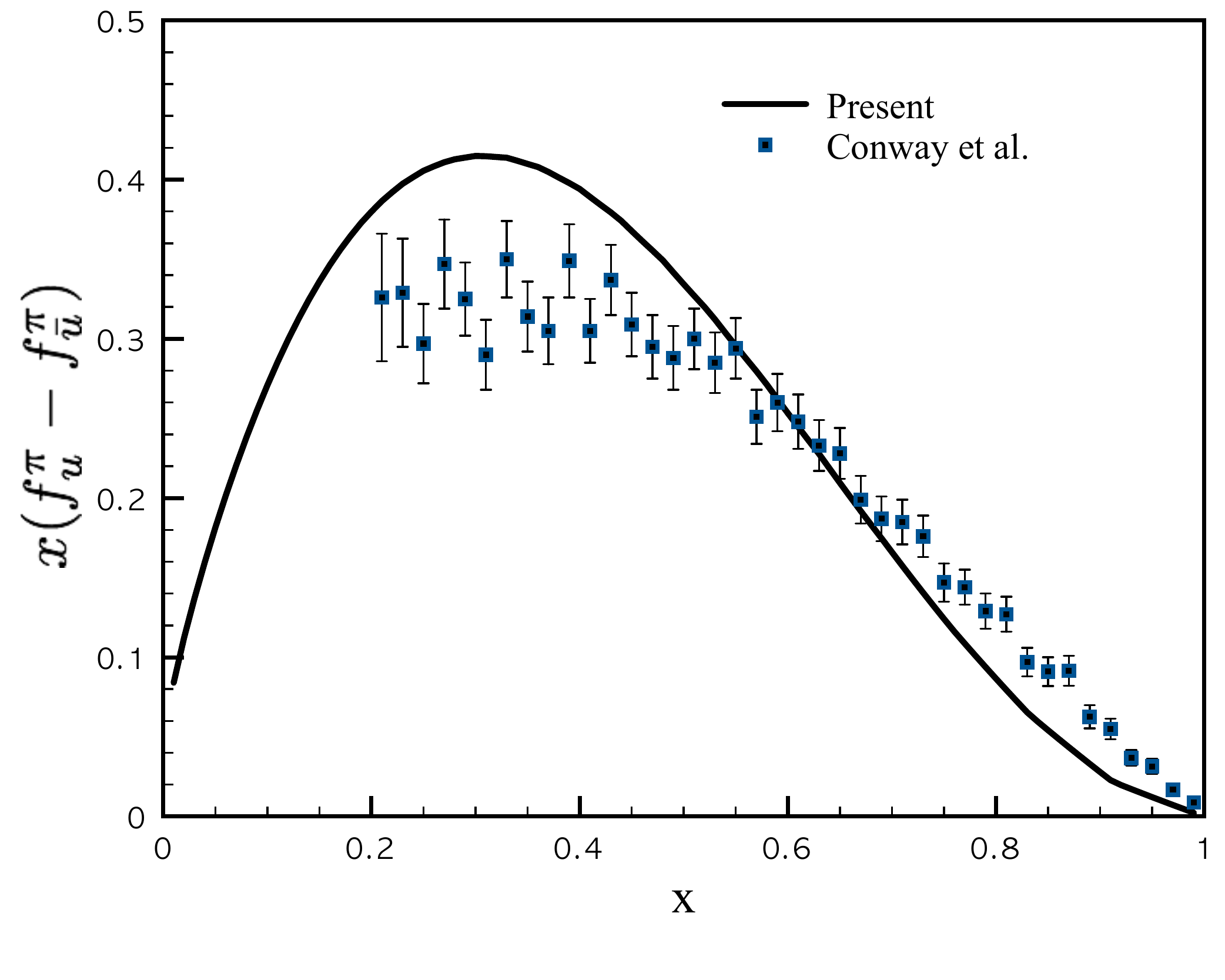}
\caption{(Color online) Valance-quark distribution function, $x(f^\pi_u-f^\pi_{\bar{u}})$  via the DGLAP evolution at LO to $Q^2=27\,\mathrm{GeV}^2$. The empirical data are taken from Ref.~\cite{Conway:1989fs}.}
\label{FIG5}
\end{figure}

\section{Summary and conclusion}
We have studied the fragmentation functions and the parton distribution functions for the positively charged pion using NLChQM motivated by the instanton vacuum configuration. We have computed them at the low-renormalization scale $\Lambda^2\approx Q^2\approx0.36\,\mathrm{GeV}^2$, then evolved them up to certain higher $Q^2$ values by the DGLAP evolution. The numerical results were compared with other models with only local interactions. Below, we list our important observations in this work:
\begin{itemize}
\item The fragmentation and parton distribution functions calculated by NLChQM manifest the effects from the nonlocal-interactions between the quarks and the PS mesons. Naturally, if we reduce the nonlocal-interaction into local ones, we can obtain the well-known results such as those from the NJL model. The nonlocal interactions make significant differences in comparison to those from the models with local interactions.
\item By comparing the first moment of the elementary fragmentation function of all isospin states, $\mathcal{N}_{\pi/u}$ from all models, the pion-cloud effects turns out to be much more pronounced in NLChQM  since the intricate nonlocal quark-PS meson interaction enhances the effects. The NLChQM results behave like a mixture of the PV and PS (NJL) schemes. It can be understood by expanding quark-PS meson coupling term $\sim M_k$ by its momentum $k$.
\item The fragmentation functions from NLChQM are relatively symmetric with respect to $z$ whereas the local interaction models give asymmetric or tilted curves. After high-$Q^2$ evolution, we find substantial deviations in comparison to the empirical data, in particular, in the small $z$ region. The inclusion of the quark-jet contribution is expected to help to resolve this discrepancy.
\item The minus-type (valance) quark distribution function reproduces the empirical data qualitatively well for various $Q^2$ values, while the plus-type one indicates sizable deviations at small $x$ region.
\end{itemize}
As a conclusion, the present nonlocal-interaction model, NLChQM provides very distinctive features, which have not been observed in usual models with local interactions, and contains interesting physical implications. We are working on the more realistic ingredients, such as the resonance and the quark-jet contributions. The result will appear elsewhere. The extension of the present theoretical framework to the flavor SU(3) octet mesons for the kaons is also under progress.

\section*{Acknowledgments}
We are grateful to J.~W.~Chen and H.~Kohyama for fruitful discussions. We also thank A.~Metz for his very helpful communications. S.i.N. is very grateful to the hospitality during his staying at National Taiwan University (NTU) with the financial support from NCTS (North) in Taiwan, where the present work was initiated. The work of C.W.K. was supported by the grant NSC 99-2112-M-033-004-MY3 from National Science Council (NSC) of Taiwan. He has also acknowledged the support of NCTS (North) of Taiwan.


\end{document}